\documentclass[british,copyright,creativecommons]{eptcs}
\usepackage[latin9]{inputenc}
\usepackage{prettyref}
\usepackage{amsmath}
\usepackage{amssymb}

\makeatletter

\providecommand{\tabularnewline}{\\}

\providecommand{\event}{G.Ciobanu, M.Koutny (Eds.): Membrane Computing and\\ Biologically Inspired Process Calculi (MeCBIC 2010)} 
\usepackage{breakurl}

\usepackage{latexsym, amssymb}

\usepackage{multicol}

\usepackage{prettyref}
\newrefformat{def}{Definition~\ref{#1}}
\newrefformat{tab}{Table~\ref{#1}}
\newrefformat{sec}{Sec.~\ref{#1}}
\newrefformat{fig}{Fig.~\ref{#1}}
\newrefformat{pro}{Proposition~\ref{#1}}
\newrefformat{alg}{Algorithm~\ref{#1}}
\newrefformat{SPiM}{(\ref{#1})}
\newrefformat{BAM}{(\ref{#1})}
\newrefformat{BA}{(\ref{#1})}
\newrefformat{B}{(\ref{#1})}
\newrefformat{BM}{(\ref{#1})}
\newrefformat{BRAM}{(\ref{#1})}
\newrefformat{BRA}{(\ref{#1})}
\newrefformat{Pi}{(\ref{#1})}
\newrefformat{SPiM}{(\ref{#1})}
\newrefformat{SPi}{(\ref{#1})}
\newrefformat{GSPi}{(\ref{#1})}

\makeatother

\usepackage{babel}

\begin{document}
\global\long\def\cons{\!::\!}

\global\long\def\decode#1{[|#1|]}

\global\long\def\encode#1{(|#1|)}

\global\long\def\gencode#1#2{\{\!\,\!|#1|\!\,\!\}}

\global\long\def\listnull{[]}

\global\long\def\substitute#1#2{\!\,_{\{#2:=#1\}}}

\global\long\def\idle{\mathbf{0}}

\global\long\def\prl{\mid}

\global\long\def\new#1{\nu#1\, }

\global\long\def\bang{\mathrm{*}}

\global\long\def\send#1#2{{\rm !}#1(#2)}

\global\long\def\ksend#1{{\rm !}#1}

\global\long\def\receive#1#2{{\rm ?}#1(#2)}

\global\long\def\kreceive#1{{\rm ?}#1}

\global\long\def\power#1{_{#1}}

\global\long\def\delay#1{\tau\power{#1}}

\global\long\def\fn#1{\mathrm{fn}(#1)}

\global\long\def\sn#1{\mathrm{sn}(#1)}

\global\long\def\bn#1{\mathrm{bn}(#1)}

\global\long\def\fnode#1{\mathrm{fi}(#1)}

\global\long\def\imply{\quad\Rightarrow\quad}

\global\long\def\reduce{\longrightarrow}

\global\long\def\transition#1{\stackrel{#1}{\reduce}}

\global\long\def\spiset{{\rm SPi}}

\global\long\def\spimset{{\rm SPiM}}

\global\long\def\transitive{^{*}}

\global\long\def\add{\oplus}

\global\long\def\remove{\ominus}

\global\long\def\totalin#1#2{{\rm In}_{#1}(#2)}

\global\long\def\totalout#1#2{{\rm Out}_{#1}(#2)}

\global\long\def\totalmix#1#2{{\rm Mix}_{#1}(#2)}

\global\long\def\totaldelay#1#2{{\rm Delay}_{#1}(#2)}

\global\long\def\activity#1#2{{\rm Act}_{#1}(#2)}

\global\long\def\inx#1{{\rm In}_{#1}}

\global\long\def\outx#1{{\rm Out}_{#1}}

\global\long\def\mixx#1{{\rm Mix}_{#1}}

\global\long\def\actx#1{{\rm Act}_{#1}}

\global\long\def\delayx#1{{\rm Delay}_{#1}}

\global\long\def\apparentx#1{a_{#1}}

\global\long\def\rateF{F}

\global\long\def\rate#1{\rateF_{#1}}

\global\long\def\apparentrate#1#2{{\rm App}_{#1}(#2)}

\global\long\def\nextchannel#1{{\rm Next}(#1)}

\global\long\def\byinduction{\stackrel{{\scriptstyle \mathrm{IH}}}{\Rightarrow}}

\global\long\def\false{\mathrm{False}}

\global\long\def\by#1{\stackrel{{\scriptstyle \mathrm{(#1)}}}{\Rightarrow}}

\global\long\def\terminates{\not\reduce}

\global\long\def\ratef#1{rate(#1)}

\global\long\def\wtrans#1{\transition{\rate{\theta},#1}}

\global\long\def\caserule#1{\mathbf{(#1)}}

\global\long\def\bylemma#1{\stackrel{{\scriptstyle \mathrm{lem\,#1}}}{\Rightarrow}}

\global\long\def\bydef#1{\stackrel{{\scriptstyle \mathrm{def\,#1}}}{\Rightarrow}}

\global\long\def\edge#1{\stackrel{#1}{\longrightarrow}}

\global\long\def\edgelabel#1{_{#1}}

\global\long\def\tuple#1{\tilde{#1}}

\global\long\def\define{\!=\!}

\global\long\def\equalspace{\quad=\quad}

\global\long\def\dotset{\mathrm{DOT}}

\global\long\def\sub#1{_{#1}}

\global\long\def\arc{\longrightarrow}

\global\long\def\larc#1{\stackrel{#1}{\longrightarrow}}

\global\long\def\inline#1{\lfloor#1\rfloor}

\global\long\def\store#1{{\rm Store}(#1)}

\global\long\def\scope#1{{\rm Scope}(#1)}

\global\long\def\substore#1{{\rm Sub}(#1)}

\global\long\def\alldot{.\;}

\global\long\def\expand#1{\lceil#1\rceil}

\global\long\def\choice{\;+\;}

\global\long\def\expansion{:}

\global\long\def\If{\;\mathbf{if}\:}

\global\long\def\with{\;\mathbf{with}\:}

\global\long\def\AND{\;\mathbf{and}\:}

\global\long\def\OR{\;\mathbf{or}\:}

\global\long\def\otherwise{\;\mathbf{otherwise}}

\global\long\def\Else{\;\mathbf{else}\:}

\global\long\def\Then{\;\mathbf{then}\:}

\global\long\def\graphcopy#1{\mathsf{Copy}(#1)}

\global\long\def\toplabel#1{^{#1}}

\global\long\def\bottomlabel#1{^{#1}}

\global\long\def\direction#1{\mathtt{#1}}

\global\long\def\sysprl{\prl}

\global\long\def\sysidle{\idle}

\global\long\def\interaction#1#2{(#1,#2)}

\global\long\def\mixedaction#1#2{#1?!#2}

\global\long\def\total#1#2{\mathrm{Tot}_{#1}(#2)}

\global\long\def\totalr#1{\mathrm{Tot}_{#1}}

\global\long\def\matches{\simeq}

\global\long\def\notmatches{\not\matches}

\global\long\def\random#1{{\rm Rand}(#1)}

\global\long\def\ind#1{^{#1}}

\global\long\def\indpair#1#2{(#1,#2)}

\global\long\def\totalrate#1{\rho(#1)}

\global\long\def\pr#1{Pr(#1)}

\global\long\def\standard#1{|#1|}

\global\long\def\NRMNM{\textrm{NM-NRM}}

\global\long\def\NMreactions{\{\textrm{non-Markovian reactions}\}}

\global\long\def\fun#1{{\it #1}}

\global\long\def\Root{root}

\global\long\def\Path{.}

\global\long\def\set#1{\widetilde{#1}}

\title{Stochastic Simulation of Process Calculi for Biology}

\author{Andrew Phillips
\institute{Microsoft Research \\ Cambridge, United Kingdom} 
  \email{andrew.phillips@microsoft.com}
\and
Matthew R. Lakin
  \institute{Microsoft Research \\ Cambridge, United Kingdom} 
  \email{v-mlakin@microsoft.com}
\and
Lo\"ic Paulev\'e
  \institute{IRCCyN, UMR CNRS 6597 \\ \'Ecole Centrale de Nantes, France}
  \email{loic.pauleve@irccyn.ec-nantes.fr}
}

\def\titlerunning{Stochastic Simulation of Process Calculi} 
\def\authorrunning{Phillips Lakin Paulev\'e}
\maketitle
\begin{abstract}
Biological systems typically involve large numbers of components
with complex, highly parallel interactions and intrinsic stochasticity.
To model this complexity, numerous programming languages based on
process calculi have been developed, many of which are expressive
enough to generate unbounded numbers of molecular species and reactions.
As a result of this expressiveness, such calculi cannot rely on standard
reaction-based simulation methods, which require fixed numbers of
species and reactions. Rather than implementing custom stochastic
simulation algorithms for each process calculus, we propose to use
a generic abstract machine that can be instantiated to a range of
process calculi and a range of reaction-based simulation algorithms.
The abstract machine functions as a just-in-time compiler, which dynamically
updates the set of possible reactions and chooses the next reaction
in an iterative cycle. In this short paper we give a brief summary
of the generic abstract machine, and show how it can be instantiated
with the stochastic simulation algorithm known as Gillespie's Direct
Method. We also discuss the wider implications of such an abstract
machine, and outline how it can be used to simulate multiple calculi
simultaneously within a common framework.
\end{abstract}

\section{Introduction}

Biological systems typically involve large numbers of components with
complex, highly parallel interactions and intrinsic stochasticity.
To model this complexity, numerous programming languages based on
process calculi have been developed, many of which are expressive
enough to generate unbounded numbers of molecular species and reactions.
Examples include variants of the stochastic pi-calculus~\cite{Priamietal01,Regevetal01,Phillips-Cardelli-CMSB07},
BlenX~\cite{Dematte08SIG}, the kappa calculus \cite{Danos07}, and
variants of the bioambient calculus~\cite{Bioambients,Phillips-ENTCS09}.
As a result of this expressiveness, such calculi cannot rely on standard
reaction-based simulation methods such as \cite{Gillespie01,Gibson00},
which require fixed numbers of species and reactions. Instead, a custom
simulation algorithm is typically developed for each calculus. The
choice of algorithm depends on the nature of the underlying biological
system, such as whether exact simulation is required \cite{Gillespie77,Gibson00},
whether certain reactions operate at different timescales \cite{Gillespie01,Tian-Burrage-04},
or whether non-Markovian reaction rates are needed \cite{Bratsun05,Pauleve-etal-CMSB10}. 

Rather than implementing custom stochastic simulation algorithms for
each process calculus, we propose to use a generic abstract machine
that can be instantiated to a range of process calculi and a range
of reaction-based simulation algorithms. The abstract machine functions
as a just-in-time compiler, which dynamically updates the set of possible
reactions and chooses the next reaction in an iterative cycle. The
abstract machine is instantiated to a particular calculus by defining
two functions: one for transforming a process of the calculus to a
set of species, and another for computing the set of possible reactions
between species. The abstract machine is instantiated to a particular
simulation algorithm by definition three functions: one for computing
the next reaction, one for computing the reaction activity from an
initial set of reactions and species populations, and a third for
updating the reaction activity as the species populations change over
time. Having a clear separation between the simulation algorithm and
the language specification allows us not only to easily instantiate
the machine to different process calculi, but also to add new simulation
algorithms that can be shared between calculi. Furthermore, the approach
could be used to dynamically integrate the simulation of multiple
process calculi simultaneously, acting as a common language runtime
for the simulation of process calculi for biology. 

In this short paper we give a brief summary of the generic abstract
machine of \cite{Pauleve-etal-CMSB10}, and show how it can be instantiated
with the stochastic simulation algorithm of \cite{Gillespie77}. We
also discuss the wider implications of such an abstract machine, and
outline how it can be used to simulate multiple calculi simultaneously
within a common framework.

\section{Summary of the Abstract Machine}

\begin{table}[!t]
\caption{\label{tab:Machine-Syntax}Syntax of the generic abstract machine,
where a term $T$ consists of the current time $t$, a species map
$S$ and a reaction map $R$. We let $\set I$ denote a multiset of
species $\{I_{1},..,I_{N}\}$ and $\set O$ denote a set of reactions.}

~

\begin{centering}
\begin{tabular}{lll}
\hline 
\textbf{T} & \textbf{syntax} & \textbf{description}\tabularnewline
\hline
$T$ & \texttt{$(t,S,R)$} & Time $t$, species map $S$, reaction map $R$\tabularnewline
$S$ & $\{I_{1}\mapsto i_{1},..,I_{N}\mapsto i_{N}\}$ & Map from a species $I$ to its population $i$,\tabularnewline
\texttt{$R$} & $\{O_{1}\mapsto A_{1},..,O_{N}\mapsto A_{N}\}$ & Map from a reaction $O$ to its activity $A$ \tabularnewline
$O$ & $(\set I,r,\set{I'})$ & Reaction with reactants $\set I$, products $\set{I'}$ and rate $r$.\tabularnewline
\hline
\end{tabular}
\par\end{centering}

~

\caption{\label{tab:Machine-Functions}Parameterised definition of the generic
abstract machine. If $\set I$ is a multiset $\{I_{1},..,I_{N}\}$
we write $\set I\oplus T$ for $I_{1}\oplus..\oplus I_{N}\oplus T$,
and $T\remove\set I$ for $T\remove I_{1}\remove..\remove I_{N}$
(the order is unimportant). We write $\fun{dom}(S)$ for the domain
of $S$. We also write $S(I)$ for the value associated with $I$
in $S$, and $S\{I\mapsto v\}$ for $S$ updated so that $v$ is associated
with $I$. }

~

\centering{}\texttt{}\begin{tabular}{rcl}
\hline 
\textbf{function} &  & \textbf{definition}\tabularnewline
\hline
$P\add T$ & $\triangleq$ & $\fun{species}(P)\add T$\tabularnewline
$I\add(t,S,R)$ & $\triangleq$ & $(t,S',R\cup R')$ if $\set{I'}=\fun{dom}(S)$; $I\notin\set{I'}$;
$S'=S\{I\mapsto1\}$; \tabularnewline
 &  & $\set O=\fun{reactions}(I,\set{I'})$; $R'=\fun{init}(\set O,(t,S',R))$\tabularnewline
$I\add(t,S,R)$ & $\triangleq$ & $(t,S',R\cup R')$ if $S(I)=i$; $S'=S\{I\mapsto i+1\}$; $R'=\fun{updates}(I,(t,S',R))$\tabularnewline
$(t,S,R)\remove I$ & $\triangleq$ & $(t,S',R\cup R')$ if $S(I)=i$; $S'=S\{I\mapsto i-1\}$; $R'=\fun{updates}(I,(t,S',R))$\tabularnewline
\hline
\end{tabular}
\end{table}
The syntax of the generic abstract machine is summarised in \prettyref{tab:Machine-Syntax},
and is based on the definitions of \cite{Pauleve-etal-CMSB10}. A
machine term $T$ is a triple $(t,S,R)$, where $t$ is the current
time, $S$ is a map from a species $I$ to its population $i$, and
$R$ is a map from a reaction $O$ to its activity $A$, which is
used to compute the next reaction. Each reaction is represented by
a tuple $(\set I,r,\set{I'})$, where $\set I$ denotes the multiset
of reactant species, $\set{I'}$ denotes the multiset of product species
and $r$ denotes the reaction rate. The structure of a term of the
abstract machine can be summarised as follows.

\begin{center}
\begin{tabular}{|c||c|c||c|c|}
\hline 
\multicolumn{5}{|c|}{Machine term $T$}\tabularnewline
\hline
\hline 
\multicolumn{1}{|c||}{Time $t$} & \multicolumn{2}{c||}{Species map $S$} & \multicolumn{2}{c|}{Reaction map $R$}\tabularnewline
\cline{2-5} 
 & Species & Population & Reaction & Activity\tabularnewline
\cline{2-5} 
 & $I_{1}$ & $i_{1}$ & $\set I_{1}\stackrel{r_{1}}{\longrightarrow}\set I_{1}'$ & $A_{1}$\tabularnewline
\cline{2-5} 
 & $\ldots$ & $\ldots$ & $\ldots$ & $\ldots$\tabularnewline
\cline{2-5} 
 & $I_{N}$ & $i_{N}$ & $\set I_{M}\stackrel{r_{M}}{\longrightarrow}\set I_{M}'$ & $A_{M}$\tabularnewline
\hline
\end{tabular}
\par\end{center}

To instantiate the abstract machine with a given process calculus,
it is sufficient to define a function $\fun{species}(P)$ for transforming
a process $P$ to a multiset of species, together with a function
$\fun{reactions}(I,\set{I'})$ for computing the set of reactions
between a new species $I$ and an existing set of species $\set{I'}$.
The syntax of species $I$ is specific to the choice of process calculus.
The $\fun{species}$ function is used to initialise the abstract machine
at the beginning of a simulation, while the $\fun{reactions}$ function
is used to update the set of possible reactions dynamically. This
allows systems with potentially unbounded numbers of species and reactions
to be simulated. 

To instantiate the abstract machine with a given simulation algorithm,
it is sufficient to define a function $\fun{next}(T)$ for choosing
the next reaction from a term $T$, a function $\fun{init}(\set O,T)$
for initialising a term with a set of reactions $\set O$, and a function
$\fun{updates}(I,T)$ for updating the reactions in a term affected
by a given species $I$. The abstract machine is then executed by
repeated application of the following rule. \emph{\[
\dfrac{(\set I,r,\set{I'}),t'=\fun{next}(t,S,R)}{t,S,R\transition{(\set I,r,\set{I'})}\set{I'}\add((t',S,R)\remove\set I)}\]
}Each time the next reaction is selected, it is executed by removing
the reactants $\set I$ from the machine term, adding the products
$\set{I'}$ and updating the current time of the machine. Corresponding
definitions for adding and removing species are summarised in \prettyref{tab:Machine-Functions}.
A process $P$ is added to a machine term $T$ by computing the multiset
of species $\{I_{1},\dots,I_{N}\}$ which correspond to $P$ and then
adding each of these species to the term. If a new species $I$ is
already present in the term then its population is incremented in
$S$ and the activity of the affected reactions is updated. If the
species is not already present in the term, its population is set
to $1$ in $S$ and new reactions for the species are computed, together
with their activity. The operation $T\remove\set I$ removes the species
$\set I$ from the machine term $T$, by decrementing the corresponding
species populations and by updating the affected reactions.

\section{Instantiating the Abstract Machine}

\begin{table}[!t]
\caption{\label{tab:Machine-Simulation}Instantiation of the generic abstract
machine with the stochastic simulation algorithm of \cite{Gillespie77}.
We write $\{E_{i}\mid C_{1};..;C_{N}\}$\emph{ }to denote the set
of elements $E_{i}$ that satisfy conditions $C_{1};..;C_{N}$. We
let $n_{1}$ and $n_{2}$ denote two random numbers from the standard
uniform distribution, $U(0,1)$. The function $\fun{propensity}(O,S)$
is defined in the main text.}

~

\centering{}\texttt{}\begin{tabular}{rcl}
\hline 
\textbf{function} &  & \textbf{definition}\tabularnewline
\hline
$\mathit{next}(t,S,R)$ & $\!\triangleq\!$ & $O_{\mu},t\!+t'$ if $a_{0}=\sum_{O_{i}\in\fun{dom}(R)}R(O_{i})$;
$t'=(\frac{1}{a_{0}})ln(\frac{1}{n_{1}})$; $\sum_{i=1}^{\mu-1}a_{i}<n_{2}a_{0}\le\sum_{i=1}^{\mu}a_{i}$ \tabularnewline
$\mathit{init}(\set O,(t,S,R))$ & $\!\triangleq\!$ & $\{O_{i}\mapsto\fun{propensity}(O_{i},S)\:\mid\: O_{i}\in\set O\}$\tabularnewline
$\mathit{updates}(I,(t,S,R))$ & $\!\triangleq\!$ & $\{O_{i}\mapsto\fun{propensity}(O_{i},S)\:\mid\: O_{i}\in\fun{dom}(R);O_{i}=(\tuple J,r,\tuple{J'});I\in\tuple J\}$\tabularnewline
\hline
\end{tabular}
\end{table}
An instantiation of the abstract machine with the stochastic simulation
algorithm of \cite{Gillespie77} is outlined in \prettyref{tab:Machine-Simulation}.
Each reaction $(\set I,r,\set{I'})$ is mapped to its propensity $a_{i}$,
which is computed by multiplying the rate of the reaction by the number
of distinct combinations of the reactants $\set I$. The function
$\fun{propensity}(O,S)$ computes the propensity of the reaction $O$
given the species map $S$ and is defined as follows, assuming that
reactions are either unary or binary and that $I_{1}$ and $I_{2}$
are distinct species. \begin{eqnarray*}
\fun{propensity}((\{I_{1}\},r,\set{I'}),S) & \triangleq & r\times S(I_{1})\\
\fun{propensity}((\{I_{1},I_{1}\},r,\set{I'}),S) & \triangleq & r\times S(I_{1})\times(S(I_{1})-1)/2\\
\fun{propensity}((\{I_{1},I_{2}\},r,\set{I'}),S) & \triangleq & r\times S(I_{1})\times S(I_{2})\end{eqnarray*}
The function $\fun{init}(\set O,T)$ computes the initial propensity
for each reaction in $\set O$, using the initial species populations
in $T$, while the function $\fun{updates}(I,T)$ updates the propensities
of all the reactions in $T$ for which $I$ is a reactant. Finally,
the function $\fun{next}(T)$ chooses a reaction from $T$ with probability
proportional to the reaction propensity, and computes the corresponding
duration of the reaction according to \cite{Gillespie77}. 

We have also instantiated the abstract machine to the Next Reaction
Method of \cite{Gibson00} and to the Non-Markovian Next Reaction
Method of \cite{Pauleve-etal-CMSB10}, by defining corresponding $\fun{init}$,
$\fun{next}$ and $\fun{updates}$ functions, as described in \cite{Pauleve-etal-CMSB10}.
We have used the abstract machine to implement the DNA Strand Displacement
(DSD) calculus for modelling DNA circuits \cite{Phillips-Cardelli-Interface09},
the Genetic Engineering of Cells (GEC) calculus for modelling of genetic
devices \cite{Pedersen-Phillips-Interface09}, and the Stochastic
Pi Machine (SPiM) calculus for general modelling of biological systems
\cite{Wang-etal-BMCSB09}, by defining appropriate $species$ and
$\fun{reactions}$ functions for each calculus. Simulators for these
three calculi are available online at \texttt{http://research.microsoft.com/dna},
\texttt{http://research.microsoft.com/gec} and \texttt{http://research.microsoft.com/spim},
respectively. Technical details of the instantiation of the generic
abstract machine with the stochastic pi-calculus and the bioambient
calculus are outlined in \cite{Pauleve-etal-CMSB10}. We are currently
developing an instantiation of the generic abstract machine to the
kappa calculus of \cite{Danos07}. Although the idea of integrating
different modelling and simulation methods within a common framework
is not a new one \cite{Ewald2010}, our approach is the first attempt
to formally define a generic framework for simulating a broad range
of process calculi with an arbitrary reaction-based simulation algorithm.

\section{A Common Simulation Framework}

The generic abstract machine can be used to simulate multiple calculi
simultaneously by assuming a separate species type $I_{L}$ for each
calculus $L$, together with an initial set of cross-calculus reactions
$\set{O_{0}}$. An example of a cross-calculus reaction is $I_{DSD}+I_{SPiM}\transition rI_{SPiM}+I'_{SPiM}$,
which takes a species of the DSD language, such as a known DNA vaccine
assembled via strand displacement, together with a species of the
SPiM language, such as a polymerase, and produces a corresponding
protein species in SPiM together with the original polymerase. The
reaction therefore enables the output of a strand displacement model
in DSD to interface with a cellular model in SPiM. For each dynamically
created species $I_{L}$ the function $\fun{reactions}(I_{L},\set{I'})$
calls the appropriate calculus-specific function $\fun{reactions_{L}}(I_{L},\set{I_{L}'}),$
where $\set{I_{L}'}$ denotes the subset of species in $\set{I'}$
that are of type $L$. This approach allows multiple calculi to interact
with each other within the same simulation environment, via a fixed
set of interface reactions. Further work is needed to formalise the
multi-language execution paradigm in more detail. 

The generic abstract machine can therefore be used to simulate a range
of existing process calculi within a common framework. By decoupling
the choice of calculus from the choice of simulation algorithm, multiple
calculi can re-use the same algorithm via a common interface, without
the need to implement custom simulation algorithms for each calculus.
In future, this could allow models to be constructed from components
written in different domain-specific languages, each designed to allow
a natural, concise encoding of that component. The components could
then interact dynamically via a common language runtime, allowing
integrated simulation of heterogeneous biological systems. 

\bibliographystyle{eptcs}
\bibliography{NM-SPiM,Scalable,Phillips}

\end{document}